\documentclass[preprint,showpacs,preprintnumbers,amsmath,amssymb,apsrev,aps,prb]{revtex4}
\usepackage{graphicx}
\usepackage{dcolumn}
\usepackage{bm}
\usepackage[latin1]{inputenc}
\newcommand{\comment}[1]{}
\newcommand\etal{\mbox{\textit{et al.}}}

\begin{document}

\title[]{Self-organization and symmetry-breaking in two-dimensional plasma turbulence}
\author{Wouter J.T. Bos{$^1$}, Salah Neffaa{$^2$} and  Kai Schneider{$^2$}}

\affiliation{$^1$ LMFA, UMR CNRS 5509, Ecole Centrale de Lyon -
  Universit\'e Claude Bernard Lyon 1 - INSA de Lyon, 69134 Ecully Cedex, France}
\affiliation{$^2$ M2P2, UMR 6181 CNRS \& CMI, Universit\'es d'Aix-Marseille, Marseille, France}

\begin{abstract}
The spontaneous self-organization of two-dimensional magnetized plasma is investigated within the framework of magnetohydrodynamics with a particular emphasis on the symmetry-breaking induced by the shape of the confining boundaries. This symmetry-breaking is quantified by the angular momentum, which is shown to be generated rapidly and spontaneously from initial conditions free from angular momentum as soon as the geometry lacks axi-symmetry. This effect is illustrated by considering circular, square and elliptical boundaries. It is shown that the generation of angular momentum in non-axisymmetric geometries can be enhanced by increasing the magnetic pressure. The effect becomes stronger at higher Reynolds numbers.  The generation of magnetic angular momentum (or angular field), previously observed at low Reynolds numbers, becomes weaker at larger Reynolds numbers.
\end{abstract}


\maketitle

\section{Introduction}

Understanding the coupling of a magnetic field with the motion of plasmas or conducting fluids is a challenging issue both from a fundamental and an applied perspective. In particular the self-organization of the velocity and magnetic fields at large scales is an intriguing phenomenon. One example is the dynamo-problem, studying the formation of a large scale magnetic field induced and amplified by fluid motion (see for example reference \cite{Monchaux2007} for recent experimental progress). Another example is large-scale spontaneous toroidal and poloidal rotation observed in fusion plasmas, an effect that is beneficial for confinement as it may suppress turbulence and radially extended structures. This effect may be related to the transition to an improved confinement state \cite{Lin2009}. The absence of this transition might jeopardize the success of the ITER \cite{ITER} project. The understanding of large-scale self-organization is therefore a key issue in different branches of physics and deserves detailed investigation. 

An academic example of self-organization is the spontaneous generation of angular momentum in two-dimensional hydrodynamic
turbulence. This phenomenon was discovered by Clercx \etal~
\cite{Clercx1998} by considering flow in a square domain. We note that
this effect was also present, but not recognized as such, in
calculations by Pointin and Lundgren \cite{Pointin1976}. In circular
domains it was observed to be absent \cite{Li1997,Schneider2005-2}. In \cite{Keetels2008} it was shown that the strength of the spin-up
can be controlled by increasing the eccentricity of an elliptic
domain. For recent reviews on the dynamics of two-dimensional
turbulence bounded by walls we refer to \cite{Heijst2006,Clercx2009} and for an
explanation of spin-up in terms of statistical mechanics to
\cite{Taylor2009,Chavanis1996}. 

In a recent work \cite{Bos2008-2}, it was shown that this effect is
enhanced in magnetohydrodynamics. The shape of the boundary which
contains a plasma may thus be very important in determining the
dynamics of close to two-dimensional plasma flow. In three dimensions,
the importance of the shape of the plasma container is far from trivial. Indeed, while in infinite cylinders plasma
can be retained in a static, quiescent state by the Lorentz force,
toroidal geometries are shown to induce non-zero velocities due to
visco-resistive effects \cite{Bates1998,Montgomery1999,Kamp2004}. These studies
concentrated on steady states in axi-symmetric geometry which 
could be qualified as two-and-a-half dimensional.  It is reasonable to expect that the same statement will be true in fully three-dimensional non-stationary MHD. That case will be studied in future work. Here we will consider the unsteady case, but in two space dimensions.

In the present work we will extend the investigation presented in
\cite{Bos2008-2}. Wall bounded two-dimensional MHD turbulence will be
studied, in which the solid boundaries are taken into account by the
penalization method \cite{Angot1999}. This method is relatively young
and has been applied to MHD turbulence only recently
\cite{Neffaa2008}, so that the present paper, in addition to its
physical relevance, also constitutes a check of the capability of the
method to model the influence of walls on high Reynolds number MHD
turbulence. We consider simulations in which the Reynolds number is
increased by approximately two orders of magnitude with respect to the
previous works \cite{Bos2008-2,Neffaa2008}. We consider three
differently shaped confining domains. In addition to the square and
circular geometry considered in the previous study we consider an
ellipse. The choice of this geometry is inspired by the work of
Keetels \etal~ \cite{Keetels2008} and this geometry has the
particularity with respect to the other two to be non-circular,
without the presence of sharp corners. The initial conditions are
completely free from angular momentum, where in \cite{Bos2008-2} a
small but non-zero initial angular momentum existed. It is shown that
the tendency to generate angular momentum becomes stronger at higher Reynolds number in the non-axisymmetric geometries, while it is absent in the circular container. Furthermore, the tendency to generate angular fields vanishes in the limit of large Reynolds numbers. An explanation is given for the vanishing of this magnetic angular momentum.

The remainder of the paper is organized as follows.
In Section~II the mathematical model, the governing equations and their numerical discretization are described.
Numerical results are presented in Section~III and finally conclusions and perspectives for future work are given
in Section~IV.

%

\section{Mathematical model of bounded MHD turbulence}

\subsection{Governing equations and boundary conditions}
Direct numerical simulation of high Reynolds number  MHD turbulence constitutes a challenge for computational physics due to the presence of a multitude of nonlinearly interacting spatial and temporal scales. Presently the most efficient method to solve homogeneous turbulence (both hydrodynamic and MHD) is by pseudo-spectral methods, using fast Fourier transforms \cite{Ishihara2009,Mininni2006-2}. The additional complexity induced by the presence of solid walls requires advanced numerical methods. Pure spectral simulations have been proposed and applied to study wall bounded MHD \cite{Shan1991}, but their prohibitive complexity for  increasing Reynolds numbers limits their application to flows with a relatively limited range of interacting degrees of freedom. 

An efficient method to compute flows in the presence of solid obstacles and walls is the volume penalization approach which was introduced by Angot \emph{et al.} \cite{Angot1999} for the Navier-Stokes equations and applied to hydrodynamic turbulence in \cite{Schneider2005-3,Schneider2005-2}. This method was extended to MHD turbulence in a recent work \cite{Neffaa2008}. Using this method, efficient pseudo-spectral solvers can be used to compute flows which contain solid walls and obstacles, which may even move in time \cite{Kolomenskiy2009}.

The governing equations are 
\begin{eqnarray}
\label{qte_mvt}
\frac{\partial{\bm{u}}}{\partial{t}}+\bm{u}\cdot\nabla\bm{u}= -\nabla p + \bm{j}
\times\bm{B} + \nu\nabla^2\bm{u} -\frac{1}{\epsilon}\chi(\bm{u} - \bm{u_0})\\
\label{eq_magnetic}
 \frac{\partial{\bm{B}}}{\partial{t}} = \nabla\times(\bm{u}\times\bm{B}) + \eta\nabla^2\bm{B}-\frac{1}{\epsilon}\chi(\bm{B} - \bm{B_0})\\
\label{DivFree}
\nabla\cdot\bm{u} = 0\\
\nabla\cdot\bm{B} = 0
\end{eqnarray}
with $\bm u$ the velocity, $\bm B$ the magnetic field, $p$ the pressure and $\bm j=\nabla\times \bm B$ the current density. Here $\nu$ and $\eta$ are respectively the kinematic viscosity and the magnetic diffusivity. The last term in the evolution equations for $\bm u$ and $\bm B$ is
the penalization term which allows to impose the solid boundary
conditions. Thus both the fluid-domain and the confining walls are
embedded in a $2\pi$-periodic square domain. We consider circular,
square and elliptic domains. For further details we refer to \cite{Neffaa2008}. 

The quantities $\bm{u_0}$ and $\bm{B_0}$ correspond to the values
imposed in the solid part of the numerical domain. Here we choose
$\bm{u_0}=\bm{0}$ and $\bm{B_0}={\bm B_\parallel}$.  
Here ${\bm B_\parallel}$ is the tangential component of ${\bm B}$ at the wall which is not being fixed at a constant value
but being re-computed at each time-step. Thus the normal component of the magnetic field vanishes at the wall, while the tangential component can freely evolve. This configuration corresponds to an electrically conducting fluid or plasma in a container with perfectly conducting walls, coated on the inside with a thin insulating layer \cite{Mininni2006}. In addition to the normal component of the magnetic field, the current density can not penetrate into the walls, a property which is automatically satisfied for two-dimensional flow since the current density only has a component perpendicular to the plane of the flow.
The mask function $\chi$ is equal to 0 inside the fluid domain (where the penalization terms thereby disappear) and equal to 1 inside the part of the domain which is considered to be a solid.  The physical idea is to model the solid part as a porous medium whose permeability $\epsilon$ tends to zero \cite{Angot1999,Schneider2005-3}. For $\epsilon$ $\rightarrow$ 0, where the obstacle is present, the velocity $\bm{u}$ tends to $\bm{u}_0$ and the magnetic field $\bm{B}$ tends to $\bm{B}_0$. The nature of the boundary condition for the velocity is thus no-slip at the wall.

\subsection{Numerical method}
In the case of two-dimensional flow (here in the $x-y$ plane) it is convenient to take the curl of eq. (\ref{qte_mvt}, \ref{eq_magnetic})
to obtain after simplification equations for the vorticity and the
current density, which become scalar valued (in the $z$-direction) and are perpendicular to the velocity and the magnetic field, respectively. The vorticity is defined by $\omega\bm{e}_z=\nabla\times\bm{u}$ and $j\bm{e}_z=\nabla\times\bm{B}$ denotes the current density. Furthermore we define the vector
potential $\bm{a} = \textrm{a}\bm{e}_z$ as $\bm{B} =
\nabla\times\bm{a}$ and the stream function $\psi$ as ${\bm u} =
\nabla^{\perp}\psi=(-\partial{\psi}/\partial{y},\partial{\psi}/\partial{x})$.
We discretize the evolution equations of vorticity and current density,
\begin{eqnarray}
\label{eqn_vort}
\frac{\partial{\omega}}{\partial{t}}+\bm{u}\cdot\nabla {\omega} = \bm{B} \cdot \nabla j + \nu\nabla^2 \omega \nonumber\\
- \frac{1}{\epsilon} \left(\nabla \times \left[ \chi(\bm{u} - \bm{u_0}) \right]\right)\cdot \bm e_z\\
\label{eq_curr}
 \frac{\partial{j}}{\partial{t}} + \nabla^2 ( \left[ \bm{u}\times\bm{B}\right]  \cdot \bm e_z ) = \eta \nabla^2 j \nonumber\\
- \frac{1}{\epsilon} \left(\nabla \times \left[ \chi(\bm{B} - \bm{B_0}) \right]\right)\cdot \bm e_z 
\end{eqnarray}
using a classical Fourier pseudo--spectral method.
Terms containing products and the penalization terms, are evaluated by the pseudospectral technique
using collocation in physical space.
To avoid aliasing errors, i.e. the production of small scales due to the nonlinear terms which are not resolved on
the grid, we de-aliase at each time step, by truncating the Fourier coefficients of $\omega$ and $j$ using
the $2/3$ rule.
For time integration we use a semi-implicit scheme of second order, an Euler-Backwards scheme for the
linear viscous term and an Adams-Bashforth scheme for the nonlinear terms, see e.g. \cite{Schneider2005-3}. 


\subsection{Initial conditions}

The main goal of the present work is the investigation of the
formation of large scale structures containing significant angular
momentum. We therefore want our initial conditions to respect two
criteria. In the first place we want them to be free from angular
momentum, in the second place we want them to be free from coherent
structures. One way to generate a zero-angular momentum initial
condition is, as described in \cite{Clercx2001},  to take an ensemble
of a large number of Gaussian vortices equally spaced. Half of the
vortices have positive circulation and the other vortices have
negative circulation. The disadvantage is that the initial condition
hereby contains coherent structures. A straightforward way to generate
an initial condition without coherent structures, is to start with
Gaussian random noise. The absence of phase correlations ensures that
no structures are present. We therefore initialize both vorticity and
current density fields with Gaussian random noise as in
\cite{Neffaa2008}. The Fourier transforms $\widehat{\omega}$ and
$\widehat{j}$, where $\widehat{\omega}(\bm{k}) =
\frac{1}{4\pi^2}\int\omega(\bm{x})
e^{-\imath{\bm{k}}\cdot{\bm{x}}}dx$, are initialized with random
phases and their amplitudes yield isotropic energy spectra of the form:
\begin{displaymath}
 E_u(k), E_B(k) \propto \frac{k}{(g + (k/k_0))^4},
\end{displaymath}
where $g=0.98$ and $k_0=\frac{3}{4}\sqrt{2}\pi$. This energy spectrum is peaked at the largest scales and follows a power law proportional to $k^{-3}$ at large wavenumbers. The energy spectra are thus the same for the magnetic field and the velocity field. The phases of the Fourier-modes are however chosen randomly and independently, so that the initial fields are different. The corresponding fields $\bm{u}$ and $\bm{B}$ are calculated from $\omega$ and $j$ using the Biot-Savart law. The fields contain vanishing cross-helicity $\int_\Omega  u_iB_i dA$, with $\Omega$ the flow domain. The so-generated fields are however, in general, not free from angular momentum. We note that this was the case in reference [\onlinecite{Bos2008-2}], in which the initial conditions contained a small amount of angular momentum. We want to avoid this in the present study in order to be able to answer to the question whether it is possible to generate angular momentum when initially none is present. 

Before describing how we achieve the generation of initial conditions free from angular momentum, let us recall the definition of angular momentum $L_u$ and angular field $L_B$, respectively,
\begin{eqnarray}\label{eq5}
L_u \, =  \, \int_\Omega \, \bm e_z \cdot (\bm r \times \bm u) \, dA 
\, & = & -2\int_\Omega \psi dA, \nonumber\\
L_B \, =  \, \int_\Omega \, \bm e_z \cdot (\bm r \times \bm B) \, dA
\, & = & 2\int_\Omega a dA,
\end{eqnarray}
where $\bm r$ is the position vector with respect to the 
center of the domain. Note that the equalities on the right hand side assume that $a$ and $\psi$ vanish at the boundary of the fluid domain. 
The angular field integral in terms of the vector potential $a$ has some significance for 'reduced' MHD \cite{Montgomery1982}. 
To obtain initial fields with $L_u=L_B=0$, we proceed as follows. We generate one set of fields ${\bm u_1,\bm B_1}$ with corresponding angular momenta $L_u^1$ and $L_B^1$ and a second set  ${\bm u_2,\bm B_2}$ with corresponding angular momenta $L_u^2$ and $L_B^2$. By linear combination of these conditions,
\begin{eqnarray}
\bm u= \bm u_1-\frac{L_u^1}{L_u^2}\bm u_2 \quad \quad \bm B= \bm B_1-\frac{L_B^1}{L_B^2}\bm B_2,
\end{eqnarray}
we get initial velocity and magnetic fields free from kinetic and angular momentum.

\section{ Numerical results}

We investigate in total $63$ computations in a square, circular and elliptic domain, the latter with an excentricity equal to $0.6$.  The mechanical Reynolds number and magnetic Reynolds number are defined, respectively, as
\begin{eqnarray}
\mathcal{R}_u=\frac{\mathcal{U}D}{\nu}\\
\mathcal{R}_B=\frac{\mathcal{U}D}{\eta}.
\end{eqnarray}
The Reynolds numbers are based on the initial root mean square
velocity $\mathcal U$, the domain size $D$ and the kinematic viscosity
$\nu$ and resistivity $\eta$. The magnetic Prandtl number $\nu/\eta$
is unity in all simulations so that both Reynolds numbers are
equal and denoted by $\mathcal R$. In the following we will therefore not distinguish between the
two Reynolds numbers. Two series of computations denoted by A and B were performed at a resolution of $512^2$ grid-points and at Reynolds numbers of the order $10^3$ and $10^4$ respectively, performing $10$ runs for each geometry for each Reynolds number. The third series, denoted by $C$ was performed at resolution $N^2=1024^2$, at Reynolds number of order $10^5$. 
The time is normalized by $D/\sqrt{2E_u(t=0)}$, $D$ being the typical lengthscale  of the fluid domain, \emph{i.e.} the sidelength of the square, the diameter of the circle and the longest cross-section of the ellipse. Parameters of the simulations are listed in table \ref{tab1}. 

\begin{figure*}
\centering
\setlength{\unitlength}{0.7\textwidth}
\includegraphics[width=1.\unitlength]{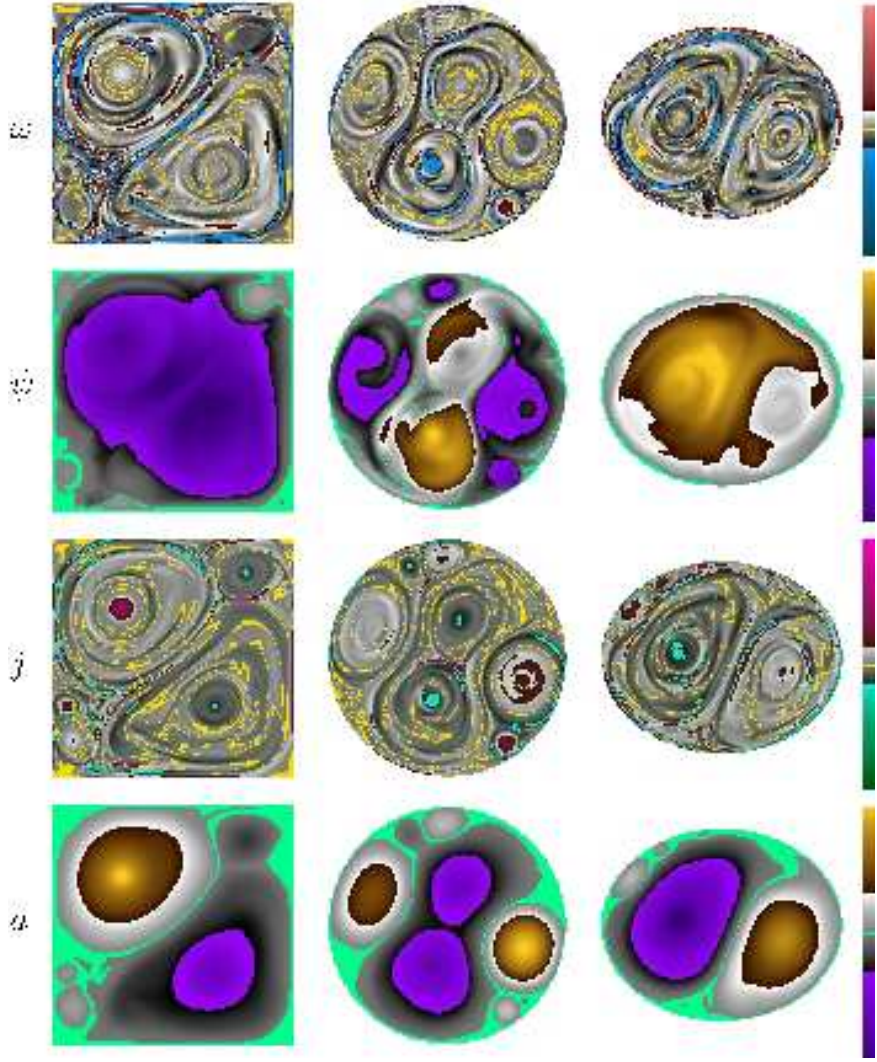}
\caption{(Color online) Visualizations of (from top to bottom) the vorticity
  $\omega$, the stream--function $\psi$, the current density $j$ and
  the vector potential $a$ for the square, circular and elliptic
  geometries. The three columns correspond
to (from left to right) to the time instants $t^\star = 3, 3, 2.7$ of series B for which $L_u$ (Fig.~2) is maximal. 
The time is normalized by the initial turn-over time. 
Note that the numerical method used in the present work does not impose a zero value of $a$ and $\psi$ at the
wall of the fluid domain. Thus a constant value was substracted from $a$ and $\psi$ at every
point in the fluid domain to impose this.}
\label{fig_flow}
\end{figure*}

\begin{table}
\caption{\label{tab1}\small{Parameters of the simulations of series A,
    B and C. $SU^*$: number of spin-up. The initial kinetic and
    magnetic energies are $E_u(0)=0.3$ and $E_b(0)=0.7$ respectively
    for all simulations. The penalization parameter $\epsilon$ is
    chosen $5\cdot 10^{-4}$ for all runs.}}
\begin{center}
\begin{tabular}{lccccc}
\hline\hline
 &  $\nu=\eta$  & $~~dt~~$ & $D$ & $SU^*$& $t^*_{max}$ \\ \hline
square (A)~~~ & $7.9\cdot10^{-4}$ & $10^{-4}$ & 2 &  1/10 & 100 \\
circle ~(A)~~~& $7.9\cdot10^{-4}$ & $10^{-4}$ & 2.24 & 0/10 & 100 \\
ellipse (A)~~~& $7.9\cdot10^{-4}$ & $10^{-4}$ & 2 & 1/10 & 100 \\ 
square (B)~~~& $1.2\cdot10^{-4}$ & $7.5\cdot10^{-5}$ & 2 & 7/10 & 100 \\
circle ~(B)~~~& $1.2\cdot10^{-4}$ & $7.5\cdot10^{-5}$ & 2.24 & 0/10 & 100 \\
ellipse (B)~~~& $1.2\cdot10^{-4}$ & $7.0\cdot10^{-5}$ & 2 & 6/10 & 100 \\ 
square (C)~~~& $1.5\cdot10^{-5}$ & $10^{-5}$ & 2 &1/1& 10\\
circle ~(C)~~~& $1.7\cdot10^{-5}$ & $10^{-5}$ & 2.24 &0/1& 10\\
ellipse (C)~~~& $1.7\cdot10^{-5}$ & $10^{-5}$ & 2 &1/1& 10\\ 
\hline\hline
\end{tabular}
\end{center}
\end{table}

\subsection{Visualizations}

Visualizations of the vorticity $\omega$, the stream-function $\psi$ the current density $j$ and the vector-potential $a$ are displayed in Figure 1. The displayed results are typical results for series B. We will first focus on the behavior in the square geometry. It is observed that both the velocity-field and the magnetic field exhibit a tendency to generate large-scale structures. The current-density shows that the magnetic field-lines of the two main flow-structures are in the opposite direction. This is even clearer in the plot of the vector potential. The magnetic angular momentum $L_B$ is therefore small, since the contributions of both structures cancel each other out. Note that the right hand side of equation (\ref{eq5}) relates the magnetic angular momentum directly to the vector potential.

In contrast, the velocity field displays significant symmetry-breaking, which is directly reflected in the stream-function. Both vortices are turning in the same sense, with a strong shearing region in between them. Non-zero angular momentum results. Similar observations can be made for the elliptic geometry. In the circular geometry it is more difficult to visually evaluate the generation of angular momentum.

\subsection{The influence of the Reynolds number and geometry}

\begin{figure}
\centering
\includegraphics[width=.48\textwidth]{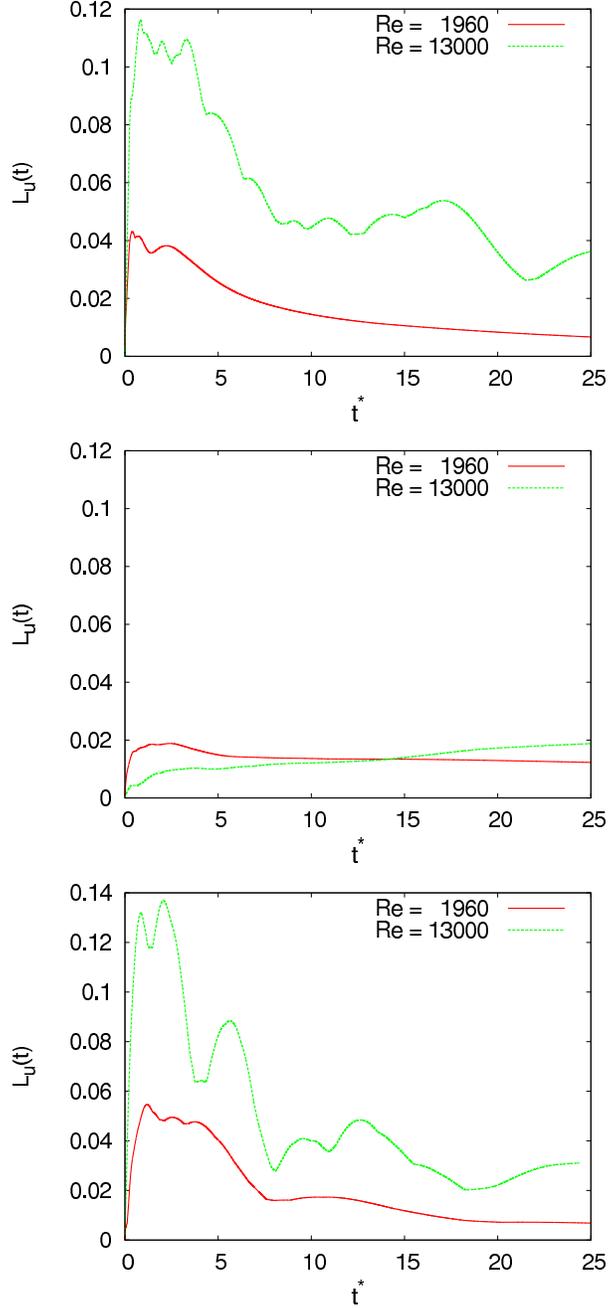}
\caption{(Color online) Influence of the Reynolds number on the spin-up: time-dependence of the absolute value of the normalized kinetic angular momentum $L_u$ averaged over 10 simulations 
of series A ($\mathcal R\approx10^3$) and series B ($\mathcal R\approx10^4$) for the
square, circular and elliptic geometry, from top to bottom. Here and
in the following the angular momentum is always normalized by
$\mathcal{L}_u(0)$ (and $\mathcal{L}_B(0)$ for the magnetic
equivalent) corresponding to the angular momentum of a solid-body having the same initial kinetic energy.}
\label{fig2}
\end{figure}

\begin{figure*}
\centering
\includegraphics[width=1\textwidth]{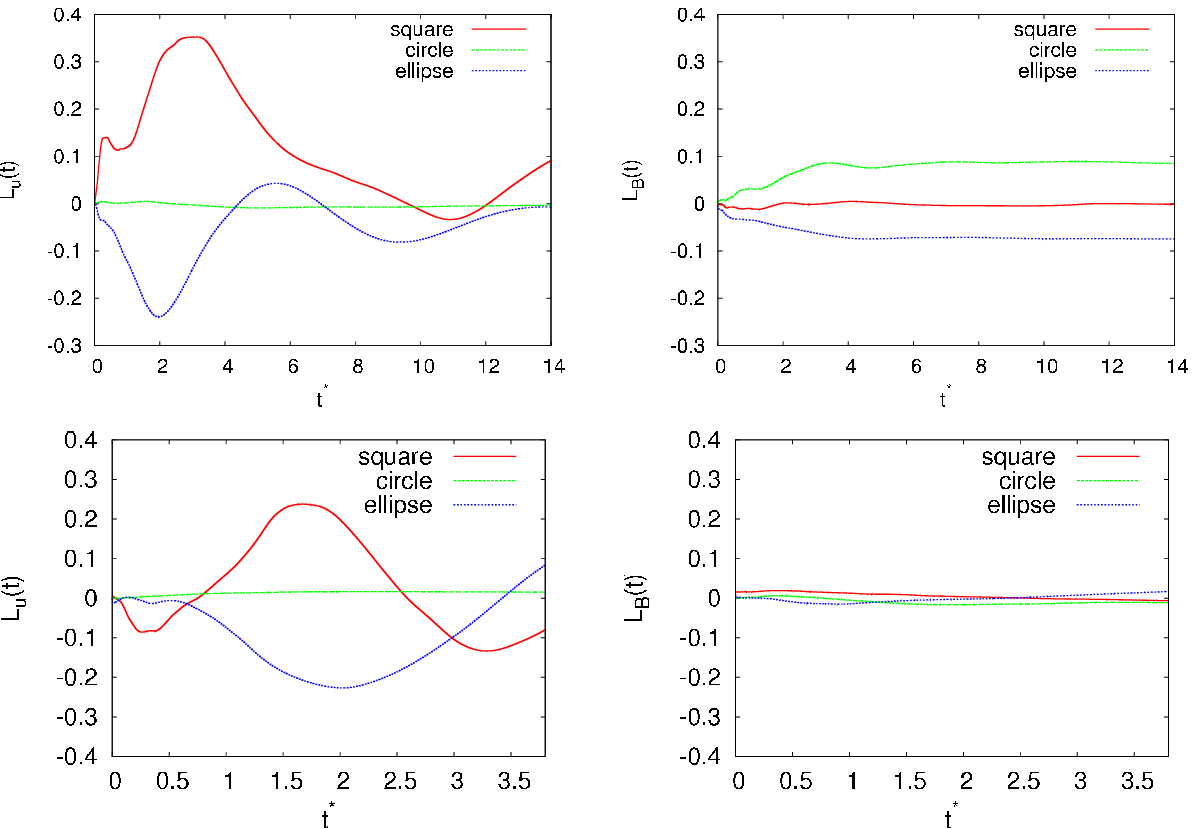}
\caption{(Color online) Comparison of series B, $\mathcal R\approx10^4$ (top) and series C, $\mathcal R\approx10^5$ (bottom). Time-evolution of the angular momentum $L_u$ (left) and angular field $L_B$ (right) in the square, circular and elliptic geometry. Only one realization is chosen from each series.}
\label{fig3}
\end{figure*}

To quantify the extent to which a large-scale swirling structure
dominates the flow, we plot in Figure \ref{fig2} the angular momentum
in the three geometries for series A and B corresponding to Reynolds
numbers of order $10^3$ and $10^4$, respectively. Since not all runs
present spin-up (a flow is defined to spin-up when the amount of
angular momentum is greater than 10\% of the angular momentum
$\mathcal L_u$ of a solid-body having the same initial kinetic energy), we show ensemble averages of the absolute value of the normalized angular momentum over ten realizations. We observe that the magnitude of the spin-up increases more than a factor 2 when increasing the Reynolds number by an order of magnitude. It is observed that the angular momentum in the circular domain is weaker but not negligible.

In Figure \ref{fig3} we show the angular momentum in the three geometries for series B and C corresponding to Reynolds numbers of order $10^4$ and $10^5$, respectively. For each Reynolds number one particular realization is chosen for which $L_u$ is maximum. For both series it is observed that strong spin-up takes place in the square and in the ellipse. The generation of the angular momentum is spontaneous, and rapid and one observes that the amplitude is of order $0.25$ in the square and in the ellipse. This implies that the fluid reaches an angular momentum which corresponds to approximately $25\%$ of the angular momentum which would possess a fluid in solid-body rotation containing the same energy at $t=0$. There is practically no spin-up in the circular container.

In Figure 3, right, the magnetic angular momentum is evaluated in all geometries. Surprisingly, in the square in which the generation of kinetic angular momentum was the strongest, $L_B$ remains close to zero. In the other two geometries, an amount of $L_B$ is created, however, this magnetic spin-up takes place on a time-scale which is larger than for its kinetic counterpart.
Furthermore it can be observed that once $L_B$ is created it remains almost constant over time. For series C $L_B$ remains close to zero at all times in all geometries.

\subsection{Influence of the magnetic pressure}  

\begin{figure}
\centering
\includegraphics[width=0.48\textwidth]{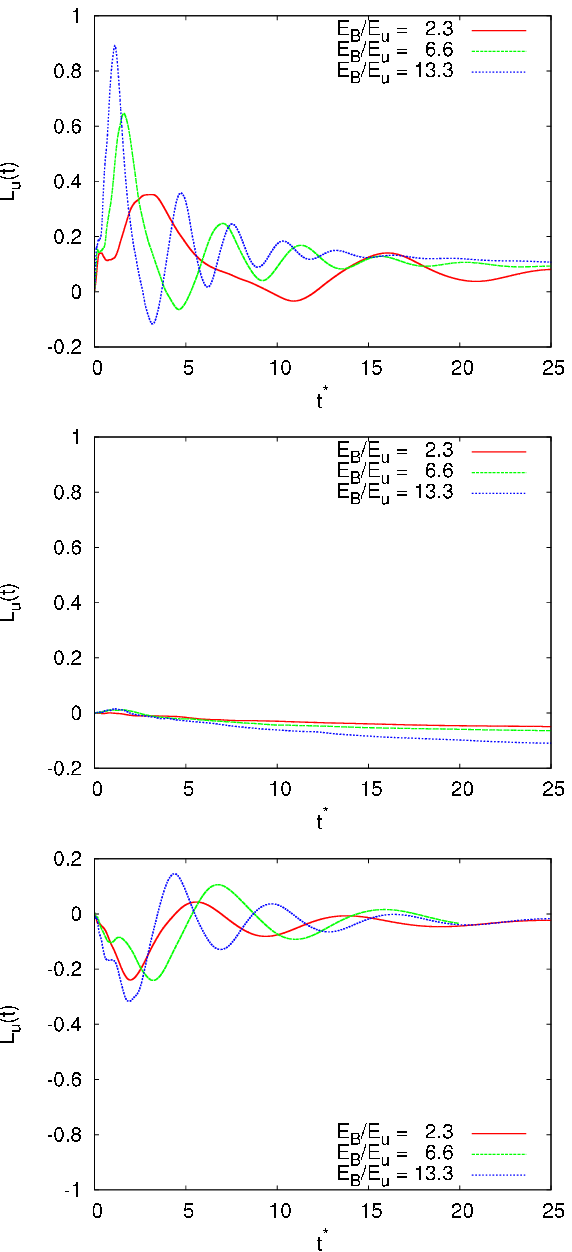}
\caption{(Color online) Time evolution of angular momentum $L_u(t)$ for series B ($\mathcal R
  \approx 10^4$). The influence of the magnetic pressure on the spin-up in the square, circle and ellipse is illustrated by changing the ratio $E_B/E_u$, while keeping $E_u$ fixed. The magnetic pressure is changed by varying $E_B$, while keeping constant $E_u$.}
\label{fig4}
\end{figure}

In \cite{Bos2008-2} we derived the equation for $L_u$ in the case of MHD turbulence. It reads
\begin{equation}\label{eq2}
\frac{d L_u}{dt} \, = \, \nu \oint_{\partial \Omega} \, \omega ({\bm r} \cdot {\bm n}) ds + \oint_{\partial \Omega} p^\star {\bm r} \cdot \, d{\bm s}
\end{equation}
with $\nu$ the kinematic viscosity, $\omega$ the vorticity, $\bm n$ the unit-vector perpendicular to the
wall, $p^\star = p + B^2/2$ is the sum of the hydrodynamic and magnetic pressure.  It was discovered by Clercx {\it et al.} \cite{Clercx1998} that spontaneous generation of angular momentum in hydrodynamic turbulence is observed
in square domains, whereas it is absent in a circular domain. Subsequently, it was
explained to be an effect due to the pressure \cite{Keetels2008}, the last term in equation (\ref{eq2}). Indeed, this term vanishes in a circular domain. In MHD, the presence of the magnetic pressure allows to vary the importance of the pressure term, while keeping the other parameters constant, by changing the value of
the magnetic fluctuations. This is illustrated in Figure \ref{fig4} for series B (Reynolds $\approx 10^4$). The ratio $E_B/E_u$ is varied,
with $E_B$ the mean-square of the magnetic fluctuations and $E_u$ the mean-square of the velocity fluctuations. It is observed that the tendency to spin-up is significantly increased in the square geometry while this effect is weaker in the elliptical geometry and absent in the circle. It is thus shown that both geometry and magnetic pressure can play a role in the generation of angular momentum. 

\subsection{On the origin of the angular fields.}

\begin{figure}
\begin{center}
\setlength{\unitlength}{\textwidth}
\includegraphics[width=.48\textwidth]{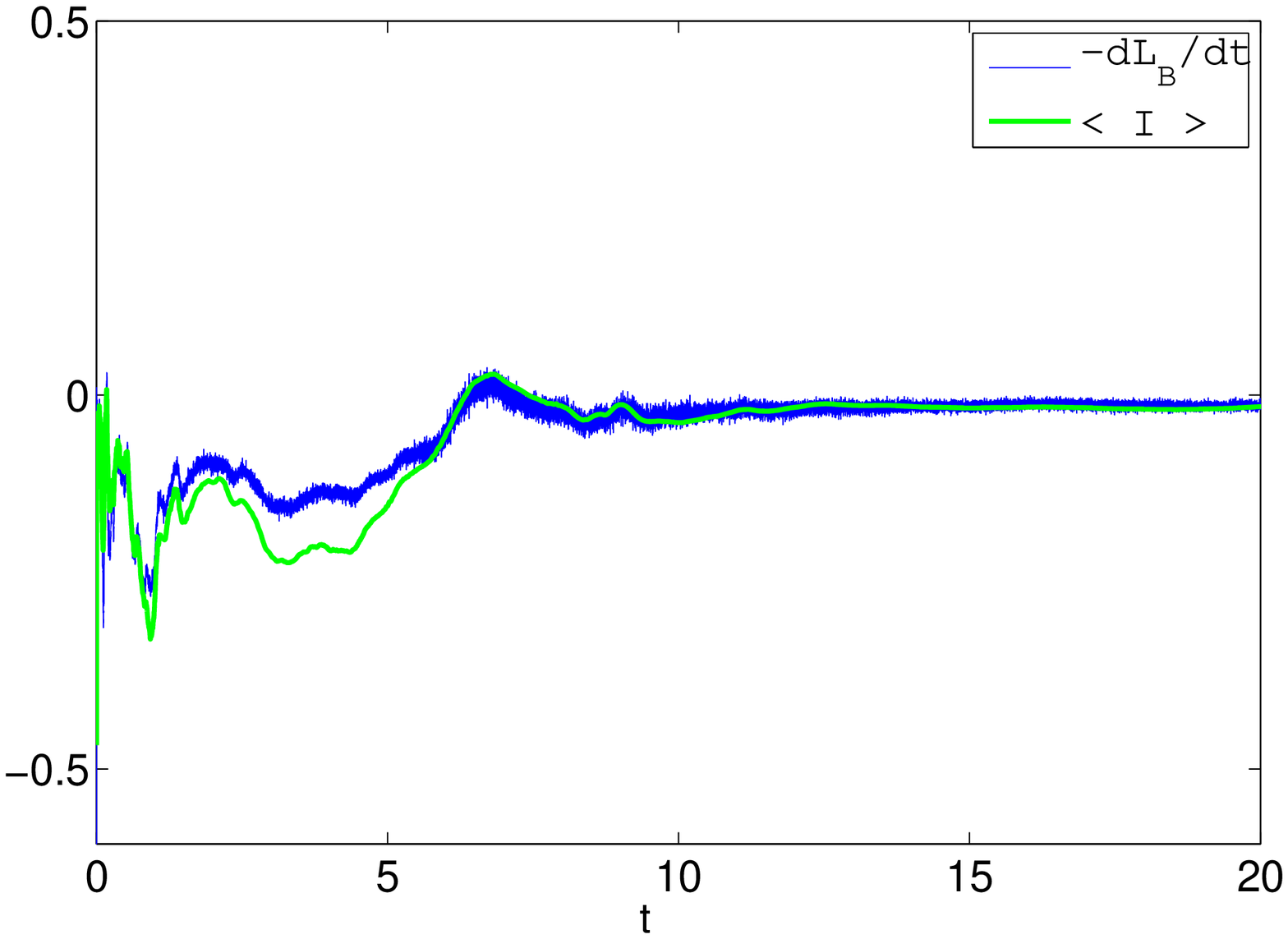}
\caption{(Color online) Comparison of the time derivative of $L_B(t)$ and the mean current $<I>$. The run corresponds to one realization in the circle with $E_B/E_u=13.3$ and $\mathcal R\sim10^4$. \label{I_dLbdt}}
\vspace{-0.5cm}
\end{center}
\end{figure}

In \cite{Bos2008-2}, the tendency to generate angular fields was also investigated by computing the value of $L_B$. It was found that angular fields were observed, even in the circular geometry. In Figure 3 right, we show that at higher Reynolds numbers the generation of this 'magnetic angular momentum' becomes weaker and seems to vanish. Writing the equation for $L_B$, we find 
\begin{equation}\label{eqdLb}
\frac{d L_B}{dt} \, = \, \eta \oint_{\partial \Omega} \, j ({\bm r} \cdot {\bm n}) ds \, - \, 2 \eta I \, ,
\end{equation}
where $I$ denotes the net current through the domain, defined by $I = \int_\Omega j dA$.
The pressure plays thus no direct role and only the net current or resistive magnetic stress can generate angular fields. 
The mean current through the domain is computed by integrating the
current density over the fluid domain. This quantity should in
principle be small, and decay to zero at long times. No production of
mean current is physically expected. Closer scrutiny of the results
revealed the existence of a spurious fluctuating mean current inside
the fluid domain. The fluctuations of this current are partly
numerical. Indeed, the penalization method is known to induce small errors in the vicinity of the wall. These errors can be controlled and depend on the parameter $\epsilon$. The thickness of the layer in which the penalization error is significant is of order $\Delta=\sqrt{\epsilon\nu}$. In this numerical boundary layer, non-physical currents can be observed. We will denote the total amount of numerical current by $I_N$. If we suppose that this current is uniformly distributed in the boundary layer, we can write for a circular domain,
\begin{eqnarray}\label{eqIm}
 I_{N} \approx 2\pi r\Delta j_{N}
\end{eqnarray}
which gives an average numerical current density $j_{N}\approx I_{N}/(2\pi r\sqrt{\epsilon\nu})$. Now, equation (\ref{eqdLb}) becomes
\begin{eqnarray}
\frac{dL_B}{dt} &\approx& r\eta 2\pi j_{N}-2\eta I_{N}\\
 &\approx& \left(r\frac{\eta}{\sqrt{\epsilon\nu}} - 2\eta \right) I_{N},
\end{eqnarray}
and for the special case of unity magnetic Prandtl number, $\nu=\eta$, this simplifies to
\begin{eqnarray}\label{eqNumI}
\frac{dL_B}{dt} &\approx& \left(r\sqrt{\frac{\nu}{\epsilon}} - 2\nu \right) I_{N}.
\end{eqnarray}
The fact that we have a penalization parameter of the order of the viscosity leads to a non-negligible production of magnetic angular momentum through the dissipation term, proportional to $I_{N}$. As one can see in Figure \ref{I_dLbdt}, the time evolution of the mean current and the time derivative of the magnetic angular momentum, computed with a classical finite difference scheme of first order, overlap quite well. Equation (\ref{eqNumI}) shows that the effect should become smaller when the ratio $\nu/\epsilon$ is decreased. Since we used the same value for $\epsilon$ in all runs and we decreased the viscosity to increase the Reynolds number, the influence of the current should become smaller at higher Reynolds number. Indeed in series C the generation of angular fields was dramatically reduced  with respect to series B as observed in Figure \ref{fig3}, which confirms our assumption that the origin is due to a numerical boundary layer. A remaining open issue is why this effect was small or absent in the square geometry. We suspect that the effect is stronger for geometries in which the mask is not aligned with the numerical grid. Indeed, a so-called \emph{staircase}-effect is expected to decrease the quality of the approximation near the walls.

\section{Conclusions and perspectives}

In total 63 pseudo-spectral simulations of two-dimensional MHD turbulence in a bounded domain were performed. It was shown that spin-up takes place in non-axisymmetric geometries (squares, ellipses). This phenomenon, observed in \cite{Bos2008-2} at low Reynolds number, persists at higher Reynolds numbers and becomes more pronounced. The generation of the magnetic equivalent of the angular momentum becomes much weaker at higher Reynolds numbers. The first effect, the kinetic spin-up can be enhanced by increasing the magnetic fluctuations. It is therefore clearly related to the pressure term $p^*$. The generation of angular fields in our simulation was shown to have a numerical origin. The effect was argued to be related to the current density leaking into the domain and can therefore be physically relevant if the walls are not assumed to be insulated. Indeed, the influence of other boundary conditions constitute an interesting objective. The main objective remains however the investigation of the effect in fully three-dimensional unsteady MHD simulations.

\section*{Acknowledgments}

David Montgomery is acknowledged for his contructive comments. We thankfully acknowledge financial support from the ANR, project M2TFP and KS
also thanks the Institut Carnot STAR for partial financial support. 

\section*{References}


\begin{thebibliography}{10}

\bibitem{Monchaux2007}
R.~Monchaux, M.~Berhanu, M.~Bourgoin, M.~Moulin, P.~Odier, J.-F. Pinton,
  R.~Volk, S.~Fauve, N.~Mordant, F.~P\'etr\'elis, A.~Chiffaudel, F.~Daviaud,
  B.~Dubrulle, C.~Gasquet, L.~Mari\'e, and F.~Ravelet,
\newblock Phys. Rev. Lett. {\bf 98}, 044502 (2007).


\bibitem{Lin2009}
Y.~Lin, J. E. Rice, S. J. Wukitch, M. J. Greenwald, A. E. Hubbard, A. Ince-Cushman, L. Lin, E. S. Marmar, M. Porkolab, M. L. Reinke, N. Tsujii and J. C. Wright,
\newblock Phys. Plasmas {\bf 16}, 056102 (2009).

\bibitem{ITER}
R. Aymar, V.A. Chuyanov, M. Huguet, Y. Shimomura and ITER Joint Central Team and ITER Home Teams,
\newblock Nucl. Fusion {\bf 41}, 1301 (2001).



\bibitem{Clercx1998}
H.~J.~H. Clercx, S.~Maassen, and G.~J.~F. van Heijst,
\newblock Phys. Rev. Lett. {\bf 80}, 5129 (1998).

\bibitem{Pointin1976}
Y.~B. Pointin and T.~S. Lundgren,
\newblock Phys. Fluids {\bf 19}, 1459 (1976).

\bibitem{Li1997}
S.~Li, D.~Montgomery, and W.~Jones,
\newblock Theor. Comput. Fluid Dyn. {\bf 9}, 167 (1997).

\bibitem{Schneider2005-2}
K.~Schneider and M.~Farge,
\newblock Phys. Rev. Lett. {\bf 95}, 244502 (2005).

\bibitem{Keetels2008}
G.~Keetels, H.~Clercx, and G.~J.~F. van Heijst,
\newblock Phys. Rev. E {\bf 78}, 036301 (2008).


\bibitem{Heijst2006}
G.~J.~F. van Heijst, H.~J.~H. Clercx and D. Molenaar
\newblock J. Fluid Mech. {\bf 554}, 411 (2006).


\bibitem{Clercx2009}
H.~J.~H. Clercx and G.~J.~F. van Heijst,
\newblock Appl. Mech. Rev. {\bf 62}, 020802 (2009).

\bibitem{Taylor2009}
J.~Taylor, M.~Borchardt, and P.~Helander,
\newblock Phys. Rev. Lett. {\bf 102}, 124505 (2009).

\bibitem{Chavanis1996}
P.~Chavanis and J.~Sommeria,
\newblock J. Fluid Mech. {\bf 314}, 267 (1996).

\bibitem{Bos2008-2}
W.J.T.~Bos, S.~Neffaa, and K.~Schneider,
\newblock Phys. Rev. Lett. {\bf 101}, 235003 (2008).

\bibitem{Bates1998}
J.W. Bates and D.C. Montgomery,
\newblock Phys. Plasmas {\bf 5}, 2649 (1998).

\bibitem{Montgomery1999}
D.~Montgomery, J.~Bates, and L.~Kamp,
\newblock Plasma Phys. Control. Fusion {\bf 41}, A507 (1999).

\bibitem{Kamp2004}
L.~Kamp and D.~Montgomery,
\newblock J. Plasma Phys. {\bf 70}, 113 (2004).

\bibitem{Angot1999}
P.~Angot, C.~Bruneau, and P.~Fabrie,
\newblock Numer. Math. {\bf 81}, 497 (1999).

\bibitem{Neffaa2008}
S.~Neffaa, W.J.T.~Bos, and K.~Schneider,
\newblock Phys. Plasmas {\bf 15}, 092304 (2008).


\bibitem{Ishihara2009}
T.~Ishihara and Y.~Kaneda,
\newblock Annu. Rev. Fluid Mech. {\bf 41}, 65 (2009).

\bibitem{Mininni2006-2}
P.~Mininni, A.~Pouquet, and D.~Montgomery.,
\newblock Phys. Rev. Lett. {\bf 97}, 244503 (2006).

\bibitem{Shan1991}
X.~Shan, D.~Montgomery, and H.~Chen,
\newblock Phys. Rev. A {\bf 44}, 6800 (1991).


\bibitem{Schneider2005-3}
K.~Schneider,
\newblock Comput. Fluids {\bf 34}, 1223 (2005).


\bibitem{Kolomenskiy2009}
D.~Kolomenskiy and K.~Schneider,
\newblock J. Comput. Phys. {\bf 228}, 5687 (2009).

\bibitem{Mininni2006}
P.~Mininni and D.~Montgomery,
\newblock Phys. Fluids {\bf 18}, 116602 (2006).

\bibitem{Clercx2001}
H.J.H.~Clercx, A.~Nielsen, D.~Torres, and E.~Coutsias,
\newblock Eur. J. Mech. B/Fluids {\bf 20}, 557 (2001).

\bibitem{Montgomery1982}
D.~Montgomery,
\newblock Phys. Scr. {\bf T2A}, 83 (1982).

\end{thebibliography}

\end{document}